\documentclass[12pt, paper]{article}
\usepackage{graphicx,lineno}
\title{
A Dynamical Systems Description of Privilege, Power and Leadership in Academia \\
\large{\it identifying barriers and pathways \\ to inclusive communities of excellent scholarship}}
\author{Kathryn V. Johnston \\  \\ Department of Astronomy, Columbia University \\ Center for Computational Astrophysics, Flatiron Institute}
\date{September 2019}

\begin{document}

\begin{titlepage}
\maketitle
 
\begin{abstract}
As the diversity of people in higher education grows, Universities are struggling to provide inclusive environments that nurture the spirit of free inquiry in the presence of these differences. At the extreme, the value of diversity is under attack as a few, vocal academics use public forums to question the innate intellectual abilities of certain demographic groups. Throughout my career as an astronomer, from graduate student, through professor to department chair, I have witnessed these struggles firsthand.
Exclusive cultures result in lost opportunities in the form of unfulfilled potential of all members of the institution --- students, administrators and faculty alike.
How to move steadily towards inclusion is an unsolved problem that hampers the advancement of knowledge itself.
As every scientist knows, problem definition is an essential feature of problem solution. 
This article draws on insights from dynamical systems descriptions of conflict developed in the social and behavioral sciences to present a model that captures the convoluted, interacting challenges that stifle progress on this problem.
This description of complexity explains the persistence of exclusive cultures and the inadequacy of quick or simple fixes. 
It also motivates the necessity of prolonged and multifaceted approaches to solutions. 
It is incumbent on our faculties to recognize the complexities in both problem and solutions, and persevere in responding to these intractable dynamics.
It is incumbent on our administrations to provide the consistent structure that supports these tasks. 
It incumbent on all of our constituents  --- students, administration and faculty --- to be cognizant of and responsive to these efforts. 
\end{abstract}

\end{titlepage}

\section{Starting Points}

\subsection{Academia in my ideal Universe}

\begin{figure}[t]
\begin{center}
\includegraphics[width=5.0in]{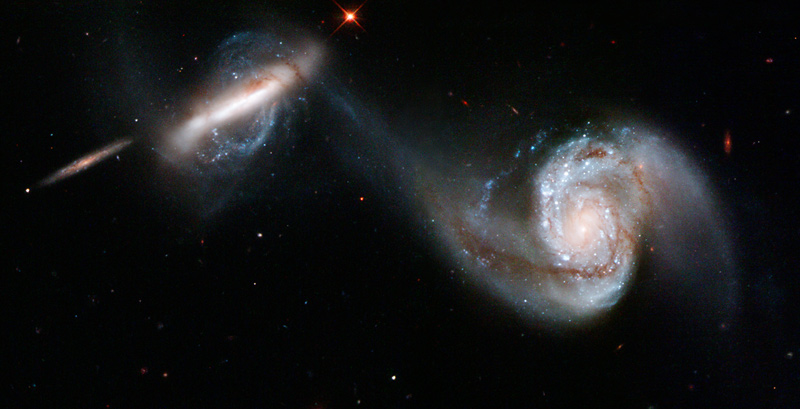}
\end{center}
\caption{\label{fig:gals}  Interacting galaxy pair Arp 87 (from {\tt hubblesite.org}, prepared for NASA under Contract NAS5-03127). Like organizations in human society, this  physical system is shaped by the mutual dynamical interactions of its constituents. These interactions are both individual (microscopic) and collective (macroscopic) in nature and are perpetually proceeding on many scales.}
\end{figure}

I am a Professor of Astronomy at Columbia University. My research explores the encounters and mergers between galaxies that are responsible for shaping these magnificent structures in the Universe, as shown in Figure \ref{fig:gals}. I teach classes on galactic dynamics, on stars and on cosmology. I mentor students and junior researchers in their own projects. And, along with my fellow faculty, I share the power and responsibility of setting the research and education mission of our great university.

It's not hard to be motivated by these roles. Academic institutions aspire to be places of unfettered exploration, where people congregate to pursue knowledge, share their ideas and learn from others. It is the people — students, administrators and faculty — who are the university, not the buildings or the campus. The faculty serve as both the creators of knowledge, as well as curators of knowledge generation and dissemination. Our charge is to make the very most of the human imagination, tenacity and capacity to explore.

Implicit in this ideal is the assumption that academia is a meritocracy that brings the most talented intellects together and supports them in constructive exchanges. And underlying this assumption is the requirement that our differences --- the full  {\it diversity}  of our experiences, learning and talents --- be explicitly recognized and celebrated so they can be used to maximize our collective impact on the boundaries of human knowledge.  Indeed, this  vision --- the pursuit of knowledge in an egalitarian, inclusive community --- drew many of us to research and education careers in the first place. 

\subsection{Academia in the real Universe}

There remain significant challenges to achieving these ideals. While my own ''Ideal Universe'' is unlikely to encapsulate every academic's vision, the underlying concerns are broadly applicable. For example, the demographics in science departments clearly do not yet reflect the talent pool. Studies by the American Institute of Physics' Statistical Research Center have shown that the fraction of PhD's awarded to African American students in physics has remained flat at a few percent for the last two decades, far below their representation in the US population \cite{aip14}. At the other end of the academic career path, science leadership remains dominated both culturally and demographically by (almost exclusively white) men. Indeed, during my last year as department chair in the 2016-2017 academic year, I was the only woman among the chairs in the Division of Science in Columbia's School of Arts and Sciences, and my peers and leaders in science research and education up the chain of command to Columbia University's president were all men. (This adds up to a total of 16 people including 8 other science department chairs, the Dean of Science, the Dean of Graduate Studies, the Dean of Columbia College, the Vice President for Arts and Sciences, the Vice President for Research, the Provost and the President.)

A vocal few attribute this lack of representation to the innate superiority of certain demographic groups in the pursuit of knowledge. This trivializes a very real problem and is a betrayal of the principles that academia is built on in the first place. It places artificial barriers to potential fulfillment for vast swaths of our talent pool. 

There are three very clear failures that need to be addressed: (i) disadvantages in background and opportunities \cite{mcintosh10}; (ii) biases in admission \cite{miller19}; and (iii) failure to make room for all people who are admitted.
This article focuses on the last of these failures --- the inability to effectively include, support and develop our full talent pool even after admission to graduate school or advancement into faculty positions and beyond to chairs or deans, provosts or presidents. This represents a significant underutilization of human resources in our institutions today. Moreover, the loss is not limited to a lack of contributions of the individuals themselves: research suggests that inclusion of diverse perspectives enhances the breadth and depth of discussions and leads to more creative solutions to problems \cite{phillips14} and that a positive climate enhances the contributions of all faculty \cite{sheridan17}.

In theory, we should be able to learn from the experience of different incoming groups as they are first admitted and subsequently advance through academia. We could use this learning to create cultures that welcome difference and accelerate the progress of other incoming groups. In practice, I have found that tensions around the differences we should be learning from limit the scope of constructive exchanges and progress to action even at the individual level. From conversations with Astronomers at institutions nationwide, it seems these problems are very common, and from the vast literature on the topic they are far from unique to our field.

As all scientists know, the starting point for solving a complex problem is to describe the problem itself  with great care \cite{spradlin12}. This article is a contribution to this aspect of problem solution. It reviews the work of our colleagues in social and behavioral sciences, and  summarizes it in a model which encapsulates the complex barriers that challenge inclusion in academia.

\section{A toy model to explore problems and solutions: dynamics of multicultural organizations}

The model I outline here is inspired by and adapted from the work of Peter Coleman, a professor at Columbia's Teachers' College, whose expertise is in finding resolutions to situations of extreme conflict. Coleman and his collaborators characterize social interactions apparent in multicultural organizations in the language of dynamical systems \cite{coleman17}. This is a language that is very familiar to astronomers who spend their lives thinking about analogously complex systems, influenced by physical processes on multiple scales. In many physical systems, microscopic effects can manifest with macroscopic consequences. In my own field of galaxy evolution, the fusion of atoms influences the evolution of stars, collective dynamics of stars shapes galaxies and galactic interactions play a key role in the formation of the structures we see in the Universe. Moreover, these interactions are mutual, affecting all galaxies involved in ways that can amplify their influence through self-reinforcing feedback loops. Merging galaxies, like the pair shown in Figure \ref{fig:gals}, are stunning examples of the power of such self-reinforcing loops: as the two galaxies approach they suffer distortions due to each other's tidal fields; these distortions convert orbital energy to internal energy so that the orbits decay; the decay of the orbits in turn enhances the tidal interaction and accelerates the merger rate.  

Prof Coleman's work was instrumental in helping me piece together my day-to-day experiences as a professor in science with what I had learned from colleagues in social science and psychology and thus to fill in the puzzle of the bigger picture: how the individual (what astronomers might term ``microscopic'') dynamics within our own communities can, despite our best intentions, collectively manifest as (``macroscopic'') barriers to admission and advancement of incoming groups that differ from those already present. Moreover, these dynamics can reinforce each other to  result in unwelcoming, unproductive work environments. 

Any community seeking to encourage diverse perspectives needs to recognize these dynamics. From my perspective as a white woman in astronomy, my academic community has already evolved from being dominated by and working effectively as a single culture (monoculture), and is transitioning through a difficult period when two or more cultures are present, but in tension with each other (multi-culture).  However, for the full potential of different groups to be jointly realized, we need to establish a productive poly-culture where the entire range of viewpoints and talents are effectively utilized. 

\clearpage
\begin{figure}[h]
\begin{center}
\includegraphics[width=4.in]{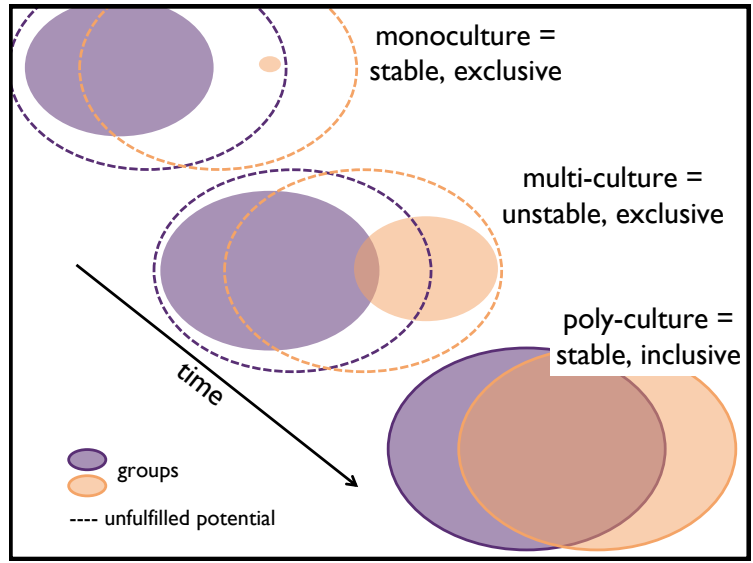}
\end{center}
\caption{\label{fig:evol} Schematic representation of in-group (purple) and out-group (orange) contributions in mono-cultural, multi-cultural and poly-cultural organizations. 
The area of the filled ellipses represents the contributions enabled by the status and influence of each group.
The dashed lines represent the potential for contributing to academia for both groups, whether through research, education or beyond. The mono-culture is dominated by one group both in numbers and in the environment it produces. In the multi-culture, more than one group may be present, yet the potential remains unfulfilled as differences are unrecognized, and hence underutilized, and the culture remains that of the in-group.  In the poly-culture, where differences are celebrated and there are open discussions of inclusive environments, both in-group and out-group members benefit.  The challenge of exploring differences leads to more robust conclusions \cite{phillips14} and the culture supports the full range of possible contributions. 
}
\end{figure}

\subsubsection{The toy model: strengths and limitations}

Figure \ref{fig:evol} provides a visual representation of our toy model that describes evolution from a monoculture (purple) to a polyculture of several groups (purple and orange). 

It is important to acknowledge that the representation in Figure \ref{fig:evol} and all subsequent plots is a vast  over-simplification. In reality there are many more than two demographic groups to be considered, and many people's experiences are complicated by sitting in the intersection of several groups \cite{prescodweinstein17}. Moreover, the linear progression shown is just one cycle of a repeating pattern as the culture in an institution is constantly evolving --- and needs to constantly evolve --- in response to the continual influx of new viewpoints. To be specific, successfully including my demographic group of white women in science will not solve the problem of including all groups. Nevertheless, our discussion will be limited to the interactions of just two groups  and one cycle of adjustment with the purpose of characterizing interactions that may be common to all such transitions. In astronomy, this would be considered a ``toy model'' of a more complex system that allows a clear and overarching vision of how to describe important facets of the evolution of such systems more generally.

In addition, the reader is warned that the toy model that is presented here, while inspired by Prof. Coleman's work, is one that reflects my own experiences in science rather than strictly reproducing the formal discussion of complex systems in social science. 

The terminology that is adopted is that used in social science unless otherwise noted. A summary of definitions is stated in Table \ref{tbl:define}, with further explanations and references given in the Appendix. The choice to use this terminology was deliberate, and again informed by experience. Being able to clearly name problems is empowering. First, it allows  individuals to realize that difficulties they may be told are their own are in fact shared by many (to the point that they are studied and named in the social sciences!). Second, naming requires careful definitions of what those problems are, which is the first step toward explicit, focussed discussions about solutions.

\begin{table}
\begin{center}
\begin{tabular}{|p {1.5 in} | p {4.in} |}
\hline
\multicolumn{2}{|c|}{Terminology} \cr
\hline
\hline
\hline
\multicolumn{2}{|c|}{usage in the article} \cr
\hline
\hline
climate & the environment of an organization as perceived by its members \cr
\hline
culture & the structures, policies, values and standards for behavior of an organization \cr
\hline
environment & the working conditions created by the culture of an organization \cr
\hline
group & people who have similar experiences of climate in an organization \cr
\hline
organization & university or other academic institution \cr
\hline
unit & department, center or institute with decision making power within an organization\cr
\hline
\hline
\multicolumn{2}{|c|}{usage in social science literature} \cr
\hline
\hline
group status threat & perceived threat to the societal status of a group \cr
\hline
homophily & the natural affinity of people for others similar to them \cr
\hline
implicit bias & unconscious attitude that may affect someone's judgment and actions \cr
\hline
imposter syndrome & concern of an individual that they have reached a position of prestige by accident \cr
\hline
masculine defaults & characteristics or behaviors that are associated with the male gender role but that are regarded as standard within an organization \cr
\hline
privilege & advantage that supports success of members of one group above another \cr
\hline
stereotype threat & negative stereotypes of a group that can lead to the underperformance of members of that group in stressful situations \cr
\hline
\end{tabular}
\end{center}
\caption{\label{tbl:define} Definitions of terms used in the article, as well as technical terms borrowed from the Social and Behavioral Science literature. For more extended discussion and references see Appendix.}
\end{table}

\subsection{The unstable multi-culture}

Most astronomy departments at universities in the United States are in the state of the unstable multi-culture. They are dominated by the western, (mostly) male group that has contributed hugely to scientific progress for several centuries, and it is this group that sets the norms (termed \underline{\it masculine defaults} in a recent article \cite{cheryan19} --- see Table \ref{tbl:define}). Many of these departments have significant fractions of (mostly white) women, but very few or zero other that are not-western or not-male \cite{aip19}.

The environment in a department can be significantly influenced by interactions at the individual, unit  (department or center within an organization) and organization (University or other academic institution) levels between the dominant (or ``in-group'') and sub-dominant (or ``out-group'') members. Many of the effects can interact as self-reinforcing and self-perpetuating feedback loops within and across the different levels.
\begin{quote}
\begin{center}
{\it \underline{\bf feedback loops across levels in an organization} \\
 individual actions and reactions manifest as collective outcomes at the unit level
\\ $\updownarrow$ \\ 
unit outcomes influence organizational policies, procedures and values  
\\ $\updownarrow$ \\ 
the culture of an organization feedbacks to structure the individuals' work environment. 
}
\end{center}
\end{quote}

\begin{figure}[t]
\begin{center}
\includegraphics[width=3.5in]{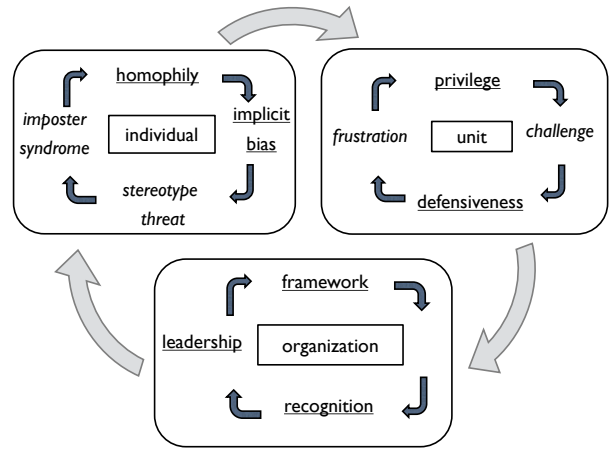}
\end{center}
\caption{\label{fig:cycles} Feedback loops in interactions at individual, unit and organization levels between in- and out-group cultures in a multi-cultural organization. Table \ref{tbl:define} outlines the usage of the terms in each box in this context. The appendix defines the terms in more depth and highlights a few relevant studies. }
\end{figure}

Figure \ref{fig:cycles} sketches how these cycles at the different levels combine to stifle progress towards inclusion, as discussed below.

\clearpage

\subsubsection{Individual dynamics}

Various terminologies, highlighted in the top-left panel of Figure \ref{fig:cycles}, have been adopted in the fields of Social Science and Psychology to isolate and identify different types of interactions at the individual level that can negatively affect the experience of the work environment. Brief definitions of these terms are given in Table \ref{tbl:define} and  expanded on with references in the Appendix. The arrows in the figure emphasize how these social interactions combine in a self-perpetuating loop that exacerbates problems.


\begin{quote}
\begin{center}
{\it \underline{\bf feedback loops in individual dynamics} \\
\underline{homophily }among members of the in-group leads to implicit bias against the out-group
\\ $\updownarrow$ \\
both homophily and \underline{implicit bias} reinforce stereotypes and trigger \underline{stereotype threat} responses in members of the out-group
\\ $\updownarrow$ \\ 
all three effects can heighten \underline{imposter syndrome} for out-group members and inhibit a sense of fitting naturally (or "belonging")
\\ $\updownarrow$ \\ 
apparent discomfort of out-group causes in-group members to withdraw, reinforcing homophily}
\end{center}
\end{quote}
This loop can exacerbate the loss of out-group members as they advance through their careers (commonly referred to as a  ``leaky pipeline'', e.g. \cite{goulden11}), lead to disengagement of those that do stay \cite{stewart18} and contributes to the lack of diverse members taking on leadership positions in academia.

\subsubsection{Unit dynamics}

The individual interactions described above can manifest collectively, and often destructively,  within a unit, as highlighted in the top-right panel of Figure \ref{fig:cycles}. 

The in-group is \underline{\it privileged}  in being able set the cultural norms that dominate a community \cite{mcintosh10}. Members of this group typically enjoy, often without recognizing that privilege, an environment that is homophilic for them, where they naturally belong and are free to concentrate on their work, unfettered by worries of implicit bias or the distracting effects of stereotype threats or imposter syndrome. 

Attempts by the out-group to broaden the culture --- through new approaches to research, teaching or community --- can be experienced as a challenge to the comfortable position of the in-group. There is increasing evidence that discomfort with these perceived challenges should be embraced as natural growing pains necessary to realize the full benefits of diversity. Discussions in diverse communities are indeed typically more contentious than homogenous ones. However, these disagreements lead to deeper and more robust explorations, and  innovative decisions \cite{phillips14}.

Unfortunately, rather than attempting to adjust and grow to encompass difference, the tendency of the in-group is defensiveness against change, which may be attributed in part to \underline{\it group status threat} \cite{wilkins13,craig14}. In addition, there exist significant social and academic inhibitions that  prevent effective discussions of difference, in particular as they approach the topics of racism, sexism and homophobia \cite{sue13}. The result is that any investigation of the origins of biased decisions can easily devolve into individuals' protesting that their intentions are not racist/sexist/homophobic and withdrawing their support for the work that needs to be done. This withdrawal pulls the in- and out-groups further apart from each other, which reinforces the impression of an exclusionary culture. The out-group reacts with frustration to this dynamic, but expression of this frustration only exacerbates the in-group's defense of both their privileged position and good intentions \cite{coleman17}.

Without explicit recognition of how individual interactions manifest destructively at the unit level, {\it both} in- and out-group members can fail to take account of the care needed to (e.g.) make unbiased decisions in order to set a positive work environment for all. In-group members in particular need to be prepared to remain engaged in uncomfortable discussions of  biases and privilege for progress towards inclusion to be made.

\vskip 0.2in
\begin{quote}
\begin{center}
{\it \underline{\bf feedback loops in unit dynamics} \\
requests for change by the out-group are experienced  by the in-group as threats to their status
\\ $\updownarrow$ \\
resistance to these discomforts by the in-group reinforces the isolation of out-group members
\\ $\updownarrow$ \\
the disconnect increases the tension between the groups
\\ $\updownarrow$ \\
this decreases their ability to engage in constructive discussions of change}
\end{center}
\end{quote}

\subsubsection{Organization dynamics}

The culture of the organization has been set historically by the in-group. Many institutions are aware of the benefits of a diverse community, and even have statements or programs to encourage hiring and retaining diverse members. However, without a consistent culture to support this breadth, decisions about reward and recognition  (i.e. promotion and tenure within the institution or the award of prizes more generally) often reflect the systematic biases at work at the individual and unit levels. These rewards reinforce the in-group norm for achievement and continue to exclude the out-group \cite{dobbin18}. Some rewards may appear merely prestigious, not carrying any formal power, but all imbue the individual with status and influence, and hence reinforce the existing culture.

The effect of lack of recognition may be exacerbated if there are  at the same time mismatches in requests for service between in-group and out-group members. Such mismatches may be a natural consequence of the gradual evolution of populations, which means that the demographics at the faculty level does not reflect  the student populations. This can lead to disproportionate mentoring and committee work falling on out-group members \cite{omeara17}. If this work is not appropriately recognized as valuable to the institution this can be discouraging to out-group members.

While many organizations also actively encourage members of out-groups to aspire to positions of prominence and leadership, there are particular challenges to leading (and learning to lead) from an out-group position. As the fraction of the out-group dwindles with distance along the pipeline, the individual dynamics of homophily, implicit bias, stereotype threat and imposter syndrome are again relevant. The standard of leadership is that of the in-group and it is this style (currently dominated by  ``masculine-defaults'' in academia \cite{cheryan19}) that is described in training programs and modeled by mentors. Yet many of the attributes of this style may be neither comfortable nor advantageous for an out-group leader to adopt. It has been shown that out-group members face increased scrutiny and criticism in leadership positions, in particular if they display behaviors that may be acceptable in the in-group but are not characteristic of expectations of the out-group \cite{stewart18}. These criticisms are exacerbated if out-group leaders are themselves advocates for diversity, with the consequence of being silenced on the very issues to which they bring special insight  \cite{hekman14,johnson16}. 

Failure to adapt to the out-group and encourage breadth in style at leadership levels both perpetuates the lack of role models that might encourage out-group members to enter academia and limits the diversity of perspectives in positions with decision-making power. The latter creates a glass ceiling of inexperience, as those who have rarely been excluded struggle to imagine what is missing in their attempts to include. This increases the difficulty of re-framing institutional culture towards inclusion.

The bottom panel of Figure \ref{fig:cycles} summarizes these dynamics at the organizational level. 

\clearpage

\begin{figure}[t]
\begin{center}
\includegraphics[width=3.5in]{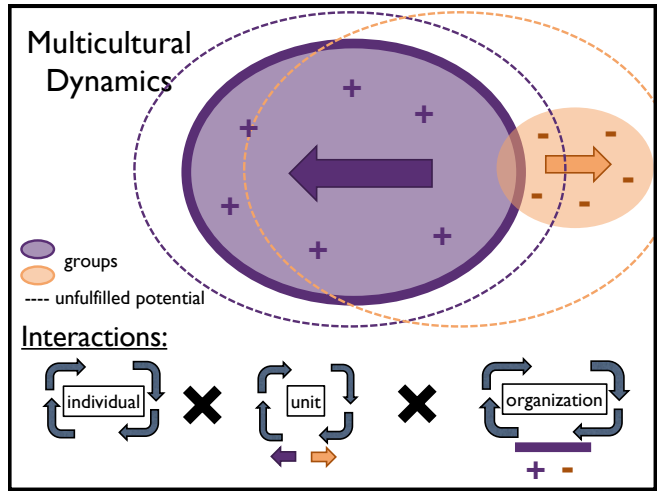}
\end{center}
\caption{\label{fig:multi} Destructive dynamics in the unstable, exclusionary multi-culture. Individual effects (see  Figure \ref{fig:cycles}) manifest collectively at the unit level as tensions between the in- and out-groups (orange and purple arrows).
The culture of the organization level reflects and supports the in-group (thick purple line). 
As a consequence, recognitions and rewards (plus signs) remain biased towards the in-group. 
Under-representation of out-group members at the faculty level relative to the student population can mean that disproportionate mentoring and committee work falls on out-group members (minus signs). 
}
\end{figure}

Figure \ref{fig:multi} summarizes how 
individual, unit and organization level effects combine. 
\begin{quote}
\begin{center}{\it \underline{\bf combining feedback loops} \\
homophily, implicit bias, stereotype threat and imposter syndrome at the individual level manifest collectively as destructive tensions between the in- and out-group (orange and purple arrows)
\\ $\updownarrow$ \\
the framework (thick purple line) at the organization level reflects and supports the in-group
\\ $\updownarrow$ \\
recognitions and rewards (plus signs) remain biased towards the in-group
\\ $\updownarrow$ \\
out-group members  overburdened with service work (minus signs)
\\ $\updownarrow$ \\
overwork and lack of recognition causes some out-group members to disengage and slows the progress of others  towards leadership positions 
\\ $\updownarrow$ \\
persistent monoculture at the leadership level stymies the ability of an organization to effectively foster diverse communities}
\end{center}
\end{quote}
In this state, neither out-group nor in-group populations are able to fulfill their potential as the vibrancy of free and open interactions is lost.

\section{Transitioning from Multi-culture to Poly-culture}

On a recent panel of under-represented minority students sharing their experiences in graduate school in STEM fields, one panelist demanded:
\begin{center}{\it Don't just tell me about imposter syndrome: \\ stop treating me like an imposter!}\end{center}
This demand eloquently characterizes our inability to include members of out-groups after they are admitted to our institutions. 

This issue of {\it Nature Astronomy} highlights some initiatives in our community that have shown to positively impact diversity efforts (e.g. \cite{yen19}). Elsewhere in the literature, there has been recent success reported in addressing gender-bias that had been found in reviews of observing proposals \cite{reid14}  by adopting a double-blind review process \cite{strolger19}. Faced with the uncomfortable evidence of systematic demographic biases in the results of the Physics GRE (summarized in \cite{miller19}), members of our community have reported correlations (or lack of correlations) with these scores for the subsequent career paths of admitted to graduate school (e.g. \cite{levesque15}), and this discussion has encouraged many departments to review their graduate-admissions process.

While the success of individual initiatives is encouraging, there is growing evidence that some common first responses to enhancing climate (e.g. missions statements, codes of conduct, diversity trainings), if adopted in isolation, can be ineffective or even counter-productive to genuine progress on inclusion \cite{dobbin18}.  The dynamical systems literature, from which the descriptive model presented here was adapted, explains these failings by elucidating the complex and subtle nature of interactions within and across individual and department levels around diversity efforts \cite{coleman17}. This literature also: clarifies why individual initiatives need to be placed within a larger, co-ordinated and multifaceted program that looks comprehensively and critically at the institutional context; and motivates finding solutions that disrupt the feedback loops that reinforce destructive patterns of interaction while emphasizing the loops that support the positive patterns. Any such comprehensive program needs to be institutionalized so that the principles of embracing and benefitting from difference can be sustained as populations entering academia continue to evolve. Research suggests the effectiveness of ``inclusive multi-culturism'' where the inclusion of {\it both} the out-group and in-group are explicitly endorsed and ``perspective training'' to support engagement across viewpoints \cite{galinksy15}. Efforts at at  Columbia can be described in these contexts: a recent survey of climate for faculty in the School of Arts and Scientists acted as a {\it disruption} when it uncovered significant gender-differences in perceptions of climate \cite{columbia18}; and we are now in the middle of building our own institutionalized and comprehensive program with a variety of approaches in response.

\begin{figure}[t]
\begin{center}
\includegraphics[width=4.5in]{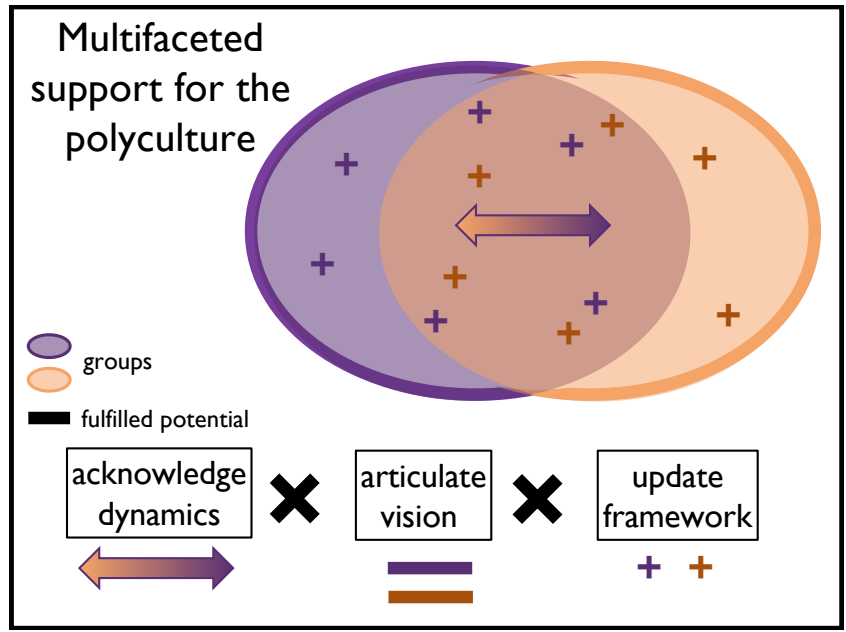}
\end{center}
\caption{\label{fig:poly} Breaking the destructive cycles described in Figures \ref{fig:cycles} and \ref{fig:multi} in the stable, inclusive poly-culture. 
Acknowledging the dynamics endorses the value of both the in- and  out-groups (double-headed arrow).  Articulating guiding principles that encourage exploration of difference promotes an environment that removes limits on potential  (thick outlines in purple and orange).  Updating the framework --- cultural norms in policies, procedures and rewards --- to include the full breadth of possible  contributions provides support for all individuals in their pursuit of  excellence (plus signs).  }
\end{figure}

Figure \ref{fig:poly} summarizes one perspective on approaches to solutions within the model above, where each component is essential and ineffective without the others.
\vskip 1.in
\begin{quote}
\begin{center}
{\it \underline{Implications: the importance of multifaceted solutions} \\
\vskip 0.1in
\underline{Acknowledging the dynamics} endorses the value of both the in- and  out-groups and allows them to confront the discomforts of culture change together (indicated schematically by double-headed arrow). 
\\ $\updownarrow$ \\
\underline{Articulating guiding principles} is itself an act of inclusion by encouraging an institution-wide culture that promotes an environment encompassing the full potential of all groups (updated outlines in purple and orange). Without being explicitly written, the unacknowledged principles remain those of the in-group. 
\\ $\updownarrow$ \\
\underline{Updating the framework} so that all policies and procedures governing admission, promotions and rewards, as well as training programs (e.g. in teaching or leadership) include all groups (``'' signs). }
\end{center}
\end{quote}
None of these will succeed without the engagement of those who build the culture of the institution ---  a partnership of \underline{faculty and  administration}.

\clearpage
\section{Privilege, Power and Leadership in Academia }

The description of the social dynamics that resist culture change highlights the importance of giving agency and self-determination to key people in an organization. With many organizations this would be the people in power at the managerial level, and indeed engagement of management on these issues and introducing them to diverse junior people are both advocated as useful \cite{dobbin18}. While this sounds like a question of leadership at the institutional level in academia, the power structure within universities is rather unique. Many relevant decisions --- the admission and advancement of graduate students, postdocs and faculty --- are made at the department level, jointly by the faculty. It is also the faculty who set the day-to-day environment for each other, their students and department staff. Yet the reward structures at many academic institutions do not recognize constructive team behaviors, but rather encourage faculty to excel as individual leaders with intense focus on and promotion of their specialized research subfield. These mixed messages can lead to an imbalance between individual and collective interests.

\vskip 0.2in
\begin{quote}
\begin{center}
\underline{\bf Conclusion: shared responsibility} \\
{\it Faculty at academic institutions --- and tenured faculty in particular --- enjoy the power and privilege of being the gatekeepers and cultivating the environment in their departments. Consequently, these individuals must each be prepared to take on the responsibilities of leadership and broad vision to encompass differences. 
\\ $+$ \\
Real change will require teamwork.
Faculty need to recognize, diagnose and openly discuss possible biases in the decisions they make, as well as address actions that promote an exclusive culture. 
They must surrender some of the comfort of familiar norms. Inclusion is not about replacement of those norms. It is about growth, enhancement and fulfillment of potential for all department members. 
\\ $+$ \\
Faculty and administration need to work together to build a framework for their Universities that rewards this teamwork in tandem with supporting individuals to pursue their research interests.
\\ $+$ \\
Research has shown that ``out-group'' leaders face unique challenges in trying to promote discussions of diversity and change, so it is crucial for in-group members to actively advocate for such considerations. }
\end{center}
\end{quote}

We need courage, resolution and tenacity to persist with this difficult and unwelcome task of self-criticism. But this persistence is imperative if academia is to live up to the ideals of inclusion that support free investigation and discovery.
\clearpage
\noindent{\bf Acknowledgements} \\
KVJ acknowledges support from NSF grant AST-1715582. She is grateful for the hospitality of the Kavli Institute for Theoretical Physics (supported by NSF grant PHY-1748958) and the University of California at Santa Cruz during her sabbatical.
She thanks her fellow members of Columbia's Committee on Equity and Diversity, colleagues in Astronomy (Suzanne Hawley, Shirley Ho, David Hogg, Jarita Holbrook, Juna Kollmeier, Alice Shapley, Rachel Somerville, David Spergel) and beyond (Peter Coleman, George Justice, Cheryl Kaiser and Diana Kardia) and friends in the PDPG for feedback as the discussion in this paper was being developed. She thanks Bridget Starling for reading and re-writing.

\clearpage
\subsection{Appendix: Glossary of effects and some associated studies}

This brief glossary is intended to expand on the effects discussed, with a few pointers to some of the research. The book ``An Inclusive Academy'' by Abigail Stewart and Virginia Valian, who are experts in diversity and inclusion, gives a comprehensive summary of findings from their own work as well as the social and psychological science literature \cite{stewart18}. Peter Coleman's article, ``Promoting Constructive Multicultural Attractors'' gives a more technical and detailed discussion of the research on brain function and social dynamics that link these effects into destructive patterns of interaction \cite{coleman17}. \\
\underline{Group Status Threat} can be activated if the characteristics of a culture are devalued in the context of a situation involving intergroup contact, in particular if this suggests that one's group will lose status and resources. This dynamic is currently being studied in the context of the political beliefs and affiliations in the US in response to changing demographics \cite{craig14,wilkins13}. \\
\underline{Homophily} refers to the natural affinity of people for others similar to them in some way, leading to a preference for (e.g.) working with similar people. To give just one illustration in academia, a study of networks of scientists at large research-intensive Universities found significant gender differences between the networks of male compared to female professors \cite{belle14}. \\
\underline{Implicit Biases} are unconscious attitudes that may affect someone's judgment and actions \cite{banaji13}. Numerous studies have revealed these biases in settings relevant to academia. Examples include: gender-bias in awards of fellowships \cite{wenneras97}, language used to describe individuals in references letters \cite{trix03} and acceptance rates of papers in prestigious journals \cite{budden08}. 
Many studies have indicated that members of both the in- and out-group display these implicit biases against the out-group. \\
\underline{Imposter Syndrome} is the concern of an individual that they have reached a position of prestige by accident and will not live up to the expectations of the position, despite evidence to the contrary. 
Imposter syndrome is something that members of both the in- ands out-group are likely to experience at some point in their career, and can have measurable negative consequences, for example, in terms of salary and job satisfaction \cite{neureiter16}.\\
\underline{Masculine Defaults} is a form of bias in which characteristics or behaviors associated with the male gender role are valued, rewarded, or regarded as necessary, standard, or normal \cite{cheryan19}. \\
\underline{Privilege} in this context refers to the advantages, some subtle but gradually accruing over time and some obvious and individually significant, that support success, recognition and promotion for the careers of members of the in-group more than members of the out-group \cite{mcintosh10}. \\
\underline{Stereotype Threat} describes the situation when negative stereotypes of a group can heighten anxieties of individuals who are members of that group in stressful situations and lead to their underperformance. The term was coined to describe the results of a study of African American undergraduates at the University of Michigan which demonstrated that they performed worse on standardized tests if reminded of their race prior to the exam \cite{steele95}. Many subsequent studies have since revealed analogous negative effects on members of groups in "threatened" situations. Examples include women underperforming in math test of  if there are only male proctors in a room,  and white men missing more hoops in a basketball contest  if they are reminded of their race prior to shooting (see \cite{steele11} for a review). These threats contribute to the under-performance of out-group members on standardized test used in admissions to our own programs, such as the Physics GRE \cite{miller19}.



\end{document}